# Surface energy calculations from Zinc blende (111)/($\bar{1}\bar{1}\bar{1}$) to Wurtzite (0001)/(000$\bar{1}$): a study of ZnO and GaN

*Jingzhao Zhang[1], Yiou Zhang[1], Kinfai Tse[1], Bei Deng[1], Hu Xu[2] and Junyi Zhu[1\*]*

[1]*Department of Physics, the Chinese University of Hong Kong, Shatin, N.T. 999077, Hong Kong*

[2]*Department of Physics, South University of Science and Technology of China, Shenzhen 518055, China*

Abstract

The accurate absolute surface energies of (0001)/(000$\bar{1}$) surfaces of wurtzite structures are crucial in determining the thin film growth mode of important energy materials. However, the surface energies still remain to be solved due to the intrinsic difficulty of calculating dangling bond energy of asymmetrically bonded surface atoms. In this study, we used a pseudo-hydrogen passivation method to estimate the dangling bond energy and calculate the polar surfaces of ZnO and GaN. The calculations were based on the pseudo chemical potentials obtained from a set of tetrahedral clusters or simple pseudo-molecules, using density functional theory approaches. And the surface energies of (0001)/(000$\bar{1}$) surfaces of wurtzite ZnO and GaN we obtained showed relatively high self-consistencies. A wedge structure calculation with a new bottom surface passivation scheme of group Ⅰ and group VII elements was also proposed and performed to show converged absolute surface energy of wurtzite ZnO polar surfaces, and the result were also compared with the above method. These calculations and comparisons may provide important insights to crystal growths of the above materials, thereby leading to significant performance enhancements of semiconductor devices.

## I. Introduction

As one of the basic quantities in surface physics, absolute surface energies play important roles in faceting, roughening, and crystal epitaxial growth. Absolute surface energies determine the crystal growth rate and equilibrium crystal shape, leading to the famous Wulff construction [1], which is generally applicable for various semiconductor materials [2-8]. Additionally, absolute surface energies have close relations with the syntheses of novel nanostructures [9-12], and novel strategies of controlling crystal growths by strain or surfactants [13-21]. One of the major problems remains to be solved in epitaxial growth is how to determine the growth mode of the hetero-epitaxial layers [22,23], which is largely determined by the absolute surface energies of the substrate materials and the epi-layers [13]. Therefore, the surface energies have important implications on qualities and performance of thin films based devices [24-26]. Since it is difficult to measure absolute surface energies in experiments [27,28], the first principle calculation becomes an important approach in determining these quantities [29-31].

[\*] jyzhu@phy.cuhk.edu.hk

Semiconductors with wurtzite (WZ) structures have broad applications in modern semiconductor industry [5,32-34], and ZnO and GaN are two representative compounds with WZ structures. ZnO has attracted considerable attentions for applications of optical devices and solar cells, due to its wide energy bandgap, large exciton binding energy (60 meV), high breakdown strength, and high saturation velocity [35]. However, the practical device application using ZnO is still undergoing inherent problems, e.g. the major difficulty in p-type doping. On the other hand, although GaN and other nitrides or nitride alloys have been widely used in blue and UV optoelectronic devices [36,37], the lacking of low cost and lattice matched substrates remains a big challenge for many group III-nitrides. The epitaxial growths of these two semiconductors are often along [0001] direction on some expensive and (or) lattice unmatched substrates (e.g. sapphire or SiC). Thus, to search for good substrate materials and optimize the crystal growth and device performance, it is necessary to first determine the absolute surface energies of polar surfaces of ZnO and GaN.

Practically, for WZ structures, it is possible to calculate surface energies of symmetric non-polar surfaces, such as ($10\bar{1}0$) and ($11\bar{2}0$) surfaces by standard slab methods based on density functional theory approaches. However, it is impossible to apply this method to polar or semi-polar surfaces because of the absence of symmetry. Several attempts were made to calculate the polar surface energies of WZ ZnO [31] and GaN [30], based on a zinc blende (ZB)/WZ hetero-junction scheme [38] and the wedge structures [39] from the ZB structure. However, there are a few problems of the wedge structure method [40]: (1) the size of the wedge structure has to be quite large to reduce edge effect; (2) pseudo-H near the edge may not be stable, affecting the accuracy of the calculation; (3) it is difficult to passivate the bottom (001) surface, leading to large errors due to the short interatomic distances of ZnO and GaN. And the estimated ZB/WZ interface energy may not be accurate. As a result, the calculated absolute surface energies may have relatively large errors, with estimated numerical convergence up to 20 meV/Å$^2$ [30]. Such large errors may significantly affect the investigations of the substrates and surfactant searching, and growth mode predictions based on these results may be problematic.

ZB based (111) or ($\bar{1}\bar{1}\bar{1}$) surfaces are adopted as reasonable approximation to simulate the c and –c planes of WZ structure in the literature [30,31]. This is because (1) the formation enthalpies of ZB and WZ ZnO (or GaN) are similar; (2) the surface atoms on the ZB (111)/($\bar{1}\bar{1}\bar{1}$) planes have the same coordination and structures up to the 2$^{nd}$ nearest neighbors of the surface atoms as the WZ (0001)/($000\bar{1}$) polar planes [30]. Also, it can be very difficult, if not impossible, to use other surfaces to approximate WZ c and –c planes. Therefore, to improve the accuracies of the WZ calculations, it's important to have a highly accurate ZB results. Recently, we proposed a relatively accurate method to calculate the absolute surface energies of ZB polar surfaces using a pseudo-H passivation approach [40]. As this method yields relatively accurate absolute surface energies of ZB (111)/($\bar{1}\bar{1}\bar{1}$) polar surfaces, it is also expected to work for WZ

(0001)/(000$\bar{1}$) polar surfaces.

In this work, we apply the aforementioned pseudo-hydrogen passivation method to calculate surface energies for (0001)/(000$\bar{1}$) surfaces of ZnO and GaN. Since local structures of the pseudo-H atoms attached on the surface are intrinsic properties determined by local electronic environment of the pseudo-H and passivated surface atoms, we can assume that the above method is also applicable to the WZ (0001)/(000$\bar{1}$) polar surfaces. We tested this assumption by comparing these results obtained from various surfaces of ZB and WZ including polar and non-polar surfaces, then checked our results with a newly developed modified wedge structure, and confirmed the reliability of our results and method.

## II. Methodology and computational details

For standard slab calculations, pseudo-H atoms are usually used to passivate the dangling bonds of the bottom surface atoms for asymmetric slabs. The pseudo-H atoms carry fractional charges to maintain charge neutrality on the bottom surfaces, as well as to stabilize the bottom surface by satisfying electron-counting-rule (ECR) [41]. To study WZ (0001) surfaces (and (000$\bar{1}$) surfaces in a similar way) based on an AB compound, we can construct a slab of the AB compound. The (0001) surface is terminated by element A, and the (000$\bar{1}$) surface is terminated by element B that is passivated by fractional-charged $H_B$. Thus, the absolute surface energy of (0001) surface is given by:

$$\sigma_{(0001)} = \frac{1}{\alpha_{(0001)}} [E^{tot}_{slab} - n_A \mu_A - n_B \mu_B - n_{H_B} \mu_{H_B} - \alpha_{(0001)} \sigma^{pass}_{bot}], \quad (1)$$

where $E_{tot}(slab)$ is the total energy of the slab with bottom surface passivated, $n_A(n_B)$ is the number of A(B) atoms in the slab, $\mu_A(\mu_B)$ is the chemical potential of A(B) atom, $\mu_{H_B}$ is the chemical potential of pseudo-H $H_B$, $\alpha_{(0001)}$ is the area of (0001) surface, and $\sigma^{pass}_{bot}$ is the surface energy of the passivated bottom surface. The chemical potentials $\mu_A$ and $\mu_B$ can vary in a range, which is constrained by the thermodynamic equilibrium between the bulk AB and bulk A(B).

$$\mu_A + \mu_B = E_{tot}(AB) = E_{tot}(A) + E_{tot}(B) + \Delta H_f(AB), \quad (2)$$

where $E_{tot}(AB)$, $E_{tot}(A)$ and $E_{tot}(B)$ are total energies of the corresponding bulk solids, and $\Delta H_f(AB)$ is the formation enthalpy of AB compound. To avoid presence of either solid A or solid B, it is required that

$$E_{tot}(A) + \Delta H_f(AB) \leq \mu_A \leq E_{tot}(A). \quad (3)$$

The left hand side and right hand side of (4) correspond to A-poor limit and A-rich limit, respectively. **Pseudo chemical potential** (PCP) $\hat{\mu}_{H_B}$ is defined as:

$$\hat{\mu}_{H_B} = \mu_{H_B} + \frac{\alpha_{(0001)}}{n_{H_B}} \sigma^{pass}_{bot} = \mu_{H_B} + [\Delta E_{int} + \Delta E_{env}], [40] \quad (4)$$

where $\mu_{H_B}$ is the 'true' chemical potential of $H_B$ atom, and the part in bracket is the binding energy between the surface atom of bottom and the pseudo-H atom. This binding energy can be further decomposed into $\Delta E_{int}$ and $\Delta E_{env}$, in which $\Delta E_{int}$ represents the intrinsic property of the surface atom, and $\Delta E_{env}$ represents the electronic environment on the surface.

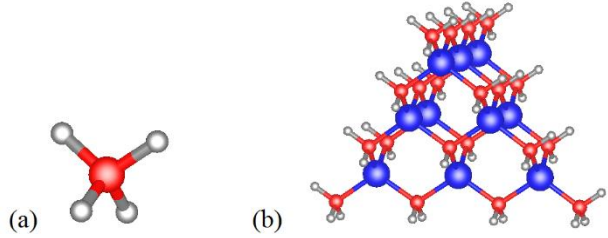

FIG. 1. Schematic illustration of the structure of a pseudo-molecule (a) and a tetrahedral cluster (b). The largest blue balls represent A atoms, the moderate red balls represent B atoms, and the smallest balls represent pseudo-H atoms $H_B$. For tetrahedral cluster in (b), number of A atoms on the edge is N = 4.

Due to the high similarity between ZB (111) surface and WZ (0001) surface, we can use the PCP of pseudo-H that passivates ZB (111) surface to estimate the PCP of the atoms that passivate WZ (0001) surface. According to our previous results in [40], to get the PCPs of pseudo-H passivated at a ZB (111) slab bottoms, we can construct either a CH4-like pseudo-molecule [**Fig. 1(a)**] or a set of tetrahedral clusters [**Fig. 1(b)**]. For the pseudo-molecule approach, the PCP is given by:

$$\hat{\mu}_{H_B} = \frac{1}{4}[E_{tot}(\text{pseudo-molecule}) - \mu_B], \tag{5}$$

where $E_{tot}$(pseudo-molecule) is the total energy of this pseudo-molecule. The tetrahedral cluster is constructed from ZB structures with four ZB (111) facets, on which all the dangling bonds are passivated by corresponding pseudo-H atoms, as shown in Fig.1 (b). The size of the cluster can be identified by N, the number of atoms on the edge. The total energy of the $N^{th}$ cluster can be expressed as:

$$E_c = \mu_A \cdot \frac{N \cdot (N+1) \cdot (N+2)}{6} + [E_{tot}(AB) - \mu_A] \cdot \frac{(N-1) \cdot N \cdot (N+1)}{6}$$
$$+ \hat{\mu}_{H_B}^{face} \cdot 2 \cdot (N-2) \cdot (N-3) + \hat{\mu}_{H_B}^{edge} \cdot 12 \cdot (N-2) + \hat{\mu}_{H_B}^{cor} \cdot 12 .. \tag{6}$$

where 4 variables are included in the expression: $E_{tot}(AB)$, $\hat{\mu}_{H_B}^{face}$, $\hat{\mu}_{H_B}^{edge}$, $\hat{\mu}_{H_B}^{cor}$. $\hat{\mu}_{H_B}^{face}$, $\hat{\mu}_{H_B}^{edge}$ and $\hat{\mu}_{H_B}^{cor}$ represent PCPs of pseudo-H atoms on the face, edge and corner of the clusters, respectively, and the total energy $E_{tot}(AB)$ of the AB compound in ZB phase is treated explicitly as one independent variable. These variables can be solved through four linear equations based on four clusters with different sizes, or by nonlinear fitting. In this work, we constructed 8 clusters with n ranging from 2 to 9 similar to previous work [40] and performed nonlinear fitting.

In addition, for polar surfaces, we can passivate both A-terminated and B-terminated surfaces of the WZ slabs to get the sum of PCPs of pseudo-H from different oriented slabs according to the following equation:

$$\hat{\mu}_{H_A} + \hat{\mu}_{H_B} = \frac{E_{tot}(\text{both}) - n_A \mu_A - n_B \mu_B}{n_H} \quad (7)$$

where $E_{tot}(\text{both})$ is the total energy of slabs with both surfaces passivated. $n_H$ is the number of pseudo-H atoms on each surface. The difference between the sum obtained from equation (7) and the sum obtained from the cluster, wedge, or pseudo-molecule method can be defined as a self-consistency check, which shows the general validity of different approaches [39].

Later, we can estimate the errors of polar surface energy calculations by using the surface energy of non-polar surfaces, which can be obtained by standard slab method. Since the Columbic interaction among nonpolar surfaces is strong, our early calculation indicates that cluster method yields relatively larger differences in the self-consistency check [40]. Therefore, it's reasonable to use the differences between the cluster method and the slab method on nonpolar surfaces to estimate the upper limit of the errors of the cluster method on the c/-c plane.

Here, a new scheme is also proposed to improve the accuracy of wedge structure algorithm, which can serve as a consistency check of the cluster method. The most crucial problem of the standard wedge structure algorithm is the problematic passivation of (001) surface of ZB structure, because the pseudo-H atoms are too close to each other, inducing significant stress on the bottom surfaces. Due to this stress, the wedge structures may have large lattice distortions, thereby decreasing the accuracies. This problem can be avoided if the (001) surfaces can satisfy ECR [41] without using pseudo-H atoms. Take ZnO as an example, for O-terminated ZB (001) surface, group-I elements may be used to passivate two O dangling bonds, each lacking $0.5e$. Similarly for Zn-terminated ZB (001) surface, we may use group-VII elements instead of the $0.5e$-charged pseudo-H atoms for the passivation. The general requirements for a good passivation element are: (1) it should not introduce gap states; (2) it should have a correct size so that little stress is induced. Our surface energy results show that, based on such modified wedge structures, the PCPs are consistent with those based on tetrahedral structures. The energies of the passivated (001) surfaces can be solved with standard symmetric (001) slabs. Then, we applied this algorithm to calculate the surface energies of ZnO and used the results as a consistency check of the cluster method.

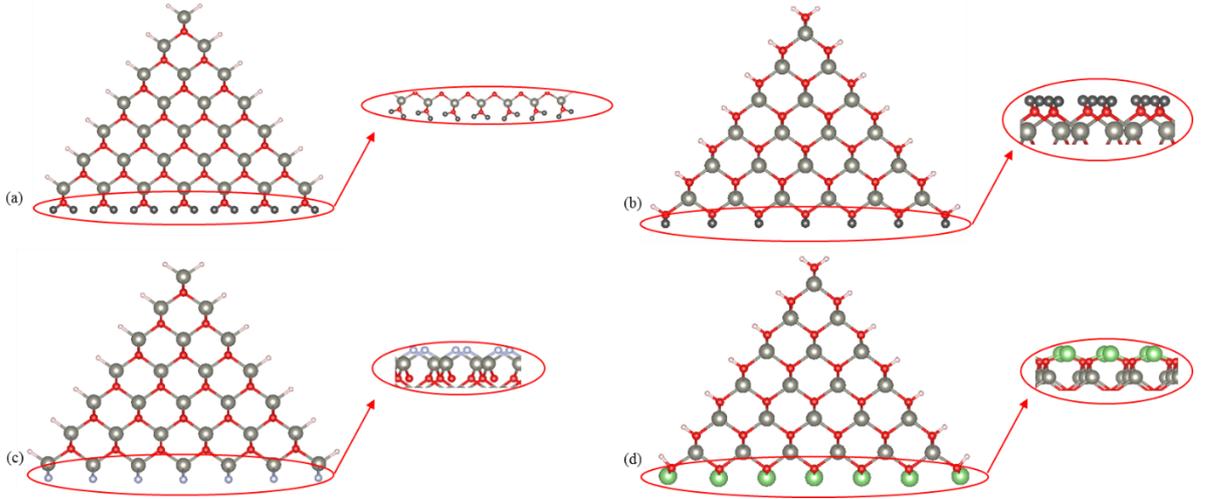

**FIG. 2**. (a) Pseudo-H passivated wedge with Zn terminated (111) surfaces. (b) Pseudo-H passivated wedge with O terminated (111) surfaces. (c) Group-Ⅶ element passivated wedge with Zn terminated (111) surfaces. (d) Group I element passivated wedge with O terminated (111) surfaces. Larger distortions induced by steric effect can be observed in pseudo H terminated wedges.

**Fig.2** shows the different passivation schemes of pseudo-H, group I and group VII elements. For instance, X is used for elements that passivate cations. And we use m to represent the sizes of wedges, i.e. the total number of zinc atoms (grey ones) shown in **Fig. 2 (b)**. For the Zn-terminated wedge in **Fig. 2 (a)**, the PCP of $H_{Zn}$ can be determined by the following equation, here taking m=36 and 28 as an example:

$$\hat{\mu}_{H_{Zn}} = (E_{wedge}^{m=36} - E_{wedge}^{m=28} - 7 \cdot \mu_O - 8 \cdot \mu_{Zn} - \hat{\mu}_X)/2 \qquad (8)$$

where $E_{wedge}^{m=36}, E_{wedge}^{m=28}$ are the total energies of two wedges of different sizes, $\mu_O$ and $\mu_{Zn}$ are the chemical potentials of O and Zn atoms. $\hat{\mu}_X$ can be calculated with a symmetric O-terminated ZB (001) slab passivated by X,

$$\hat{\mu}_X = (E_{(001)}^X - n_O \cdot \mu_O - n_{Zn} \cdot \mu_{Zn}) / n_X \qquad (9)$$

where $E_{(001)}^X$ is the total energy of the X passivated symmetric ZB (001) slab, $n_O$ and $n_{Zn}$ are the number of O and Zn atoms in the slab, $n_X$ is the number of X atoms. PCP of $H_O$ can be determined in a similar way as the corresponding O-terminated wedge structure, shown in **Fig. 2 (d)**.

The total energy calculations of bulks, slabs and clusters were based on Density Functional Theory [42,43] as implemented in VASP code [44], with a plane wave basis set [45,46] and PBE Generalized Gradient Approximation (GGA) as the exchange-correlation functional [47]. Ga d-electrons are included as valence electrons. Since GGA functional underestimates the band gap of GaN and ZnO, which may affect the energy of the surface states within the gap [30], we also performed calculations using hybrid functional of Heyd, Scuseria, and Ernzerhof (HSE) [48,49] on slabs of polar surfaces and pseudo-molecules of ZnO and GaN. Also, our method is expected to be independent of the functional. All the standard slab calculations were performed on 1×1 slabs containing 10 bilayers. A Gamma-centered k-point mesh was used for integration over Brillouin zone. The mesh is 15×15×1 for GGA calculations and 7×7×1 for hybrid functional calculations. Pseudo-molecules and clusters were calculated with Gamma point only, while modified wedge structures were calculated with 1×1×7 Gamma-centered k-point mesh. Oxygen molecule is calculated with spin polarization. The slabs and clusters were separated by at least 15Å vacuum. All the atoms in the slab and cluster were allowed to relax until forces converged to less than 0.005eV/Å for GGA calculations and 0.01 eV/Å for HSE calculations. The energy cutoff of the plane-wave basis set was set to 450eV for ZnO and 500eV for GaN for GGA, and 400eV for both ZnO and GaN for HSE. All the atoms are relaxed for all of the calculations.

## III. Results and discussion

**Table 1**

Calculated and experimental lattice constants for WZ ZnO and GaN.

| WZ | | GGA | HSE | Experiments |
|---|---|---|---|---|
| **ZnO** | **a (Å)** | 3.280 | 3.228 | 3.250 [50] |
| | **c/a** | 1.613 | 1.606 | 1.602 [50] |
| **GaN** | **a (Å)** | 3.218 | 3.163 | 3.189 [51] |
| | **c/a** | 1.629 | 1.627 | 1.626 [51] |

**Table 1** shows the lattice constants of ZnO and GaN, obtained from our bulk calculations with both GGA and HSE functionals. All these parameters are in good agreement with the experimental values and the results of HSE functional show better agreement than that of GGA functional do.

Next, we calculated the PCP of pseudo-H, based on the tetrahedral cluster method. By solving the linear equation set of the four tetrahedral clusters of different sizes, we obtained the PCPs of the pseudo-H atoms. Different selection of clusters may result in slightly different values of PCPs, which should converge when clusters are large enough. Then we tried another numerical scheme by applying a polynomial fitting method involving all 8 tetrahedral clusters to calculate the PCPs, as shown in **Fig. 3**. The results from the fitting method are very close to those obtained from linear equation set method, indicating that the obtained PCPs have converged. The fitting results of PCPs are listed in **Table 2**.

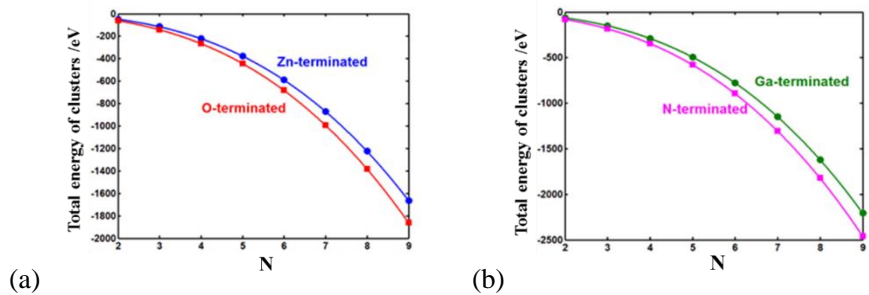

**FIG. 3.** A 3rd order polynomial fitting of the 8 clusters of ZnO (a) and GaN (b) respectively. X-terminated indicates that the atom X is located at the positions of the red atoms in Fig.1 (b).

To verify the accuracy of the cluster method, we also developed a new passivation scheme of the wedge method for ZnO. In our new scheme, Li, Na, and K are adopted to passivate O-terminated surfaces, whereas F and Cl are adopted to passivate Zn-terminated surfaces. Our calculations show that by correctly choosing the passivating elements (Li for O-terminated surface and F for Zn-terminated surface), the modified wedge method, as shown in Fig. 1 (c) and (d), can give results consistent with our tetrahedral cluster method, and the distortions of the wedge structures can be significantly reduced, as shown in **Fig. 2**. Convergence tests were performed on different O-terminated wedge structures from n=6 to n=55, as shown in **Fig. 4**. The converged PCP results were observed even for the smallest ones, and here we listed the results of n=28, 36 in Table 2. The wedges of the same sizes were also used in reference [30]. Convergence was also tested and verified on Zn-terminated wedge structures and the results are also listed in **Table 2**. By using the wedge structures shown in **Fig. 2 (b)**, we also got a relatively accurate value of PCP of $H_O$, which can be comparable with that by using structures in **Fig. 2 (d)** or cluster method. This is because that, although the structures in **Fig. 2 (b)** also have the steric effect, the steric direction is confined by periodic boundary condition, and it can be reproduced on the ZB (001) surface. In this case, there exists a large error canceling effect. However, such effect does not exist for structures in **Fig. 2 (a)**.

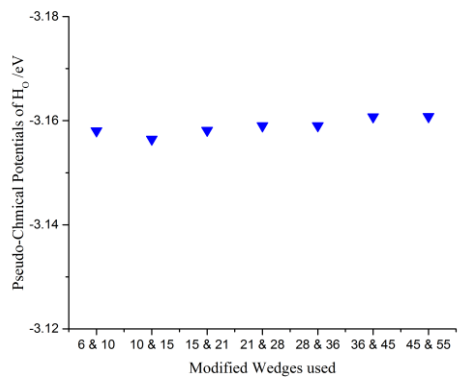

**FIG. 4.** Convergence test of calculated PCPs for $H_O$ by modified wedges. The horizontal axis means the wedges used to get the corresponding values.

**Table 2**

Sum of PCPs (eV) from ZB/WZ polar surfaces, clusters fitting and pseudo-molecules. Values in the 2$^{nd}$ line are the differences with the sum from WZ (0001)/(000$\bar{1}$) surfaces. Values in the column of "Equations solving" are the values obtained from solving equations with cluster size n = 2,3,8,9. And the sizes of the modified wedges are those of m=28,36.

|  | ZB (111)/($\bar{1}\bar{1}\bar{1}$) | WZ c/-c | Clusters Fitting | Equations solving | Modified wedges | Pseudo molecules |
|---|---|---|---|---|---|---|
| **ZnO** | -5.453 | -5.434 | -5.441[*] | -5.432 | -5.446[**] | -5.476 |
|  | -0.019 | 0.000 | -0.007 | 0.002 | -0.012 | -0.046 |
| **GaN** | -6.701 | -6.680 | -6.688[***] | -6.688 |  | -6.734 |
|  | -0.021 | 0.000 | -0.008 | -0.008 |  | -0.054 |

\*      The PCPs for $H_{Zn}$ and $H_O$ are -2.285 eV, -3.156 eV;

\*\*     The PCPs for $H_{Zn}$ and $H_O$ are -2.287 eV, -3.159 eV;

\*\*\*    The PCPs for $H_{Ga}$ and $H_N$ are -3.194 eV, -3.494 eV.

Sums of PCPs for pseudo-H are summarized in **Table 2**. These sums obtained from different approaches were compared with the values obtained from slab calculations to determine self-consistencies, which is a standard treatment in the analysis of early wedge structure calculations [39]. For $H_N$ and $H_O$, differences in PCPs between the pseudo-molecule method and cluster method are quite small, while this is not the case for $H_{Ga}$ and $H_{Zn}$. This indicates that the PCPs of $H_N$ and $H_O$ do not change significantly at different local coordinations. This is probably because O and N have very strong electronegativity, and the local electronic environment does not change much no matter they are attached to cations or to pseudo-H atoms.

Additionally, we listed the sum of PCP based on the slabs of ZB, the slabs of WZ, and pseudo molecules in Table 2. From results in **Table 2**, it can be seen that the sums obtained from polynomial fitting of these tetrahedral clusters have reasonable agreements with the sums obtained from ZB (111)/($\bar{1}\bar{1}\bar{1}$) surfaces or from WZ (0001)/(000$\bar{1}$) surfaces. The small differences between the sums obtained from ZB polar surfaces and those obtained from WZ polar surfaces indicate that it is a reasonable approximation to use PCPs obtained from clusters of ZB structures to simulate WZ (0001)/(000$\bar{1}$) polar surfaces. The dangling bond energy of the surface atom is dependent upon both intrinsic atomic orbits and the electronic environment on that surface. And the results from cluster fitting and equations solving are close to each other. Although the tetrahedral clusters are based on ZB structures, the PCP sums obtained from WZ polar surfaces are even closer to the sums than those obtained from ZB polar surfaces. This may be due to the fact that the clusters with finite sizes do not reproduce the whole $T_d$ symmetry of ZB structures. Instead, they preserve $C_{3v}$ symmetry. As a result, the clusters method turns out to be a good simulation to WZ structure, while the pseudo-molecule approach has somewhat larger errors because it only captures the physical essence of the isolated atomic orbit energy contributing to the dangling bond energy. However, the resulting accuracies of surface energies are still acceptable (in the order of several $meV/Å^2$). Comparing the values from modified wedges and clusters, we can easily find that the accuracies of them are comparable, and the differences in the absolute surface energies are within 0.3 meV/Å². Therefore, both approaches are capable of obtaining accurate surface energies for polar surfaces. Yet, tetrahedral cluster structures method yields better self- consistency.

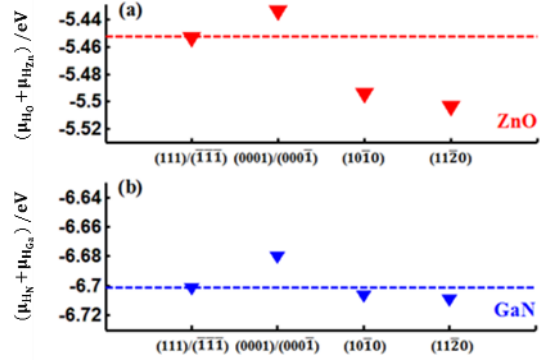

**FIG. 5.** Comparisons of sums of PCPs for different surfaces of ZB ZnO (a) and WZ GaN (b). The dashed line shows the values from ZB **(111)/($\overline{111}$)** surfaces.

**Fig. 5** shows the comparisons of the sums of PCPs for different surfaces of ZnO and GaN respectively. From these comparisons, we found that: (1) the differences among these surfaces are originated from their different electronic environments; (2) the pseudo-H bonding on nonpolar surfaces is stronger than that on polar surface; (3) the differences between polar and nonpolar surfaces in GaN are relatively smaller than those in ZnO. Our previous comparison of ZnS and GaP also showed the similar relationship [40]. The physical origin of this observation is probably due to the ionic nature of II-VI compounds, which may have strong Columbic attractions on the non-polar surfaces to stabilize the pseudo-H atoms.

**Table 3**

Summary of surface energies (meV/Å²) on different surfaces. The calculations are performed under **anion-rich limit**, i.e. $\mu_B = \mu_{B_2}/2$. All of these are bare surface energies that have been fully relaxed. ZB-(111) and WZ-(0001) means cation-terminated surfaces and ZB-($\overline{1}\overline{1}\overline{1}$) and WZ-(000$\overline{1}$) denote anion-terminated ones. Values in parentheses are surface energies calculated from standard slab calculations. And the Cluster Method is based on the fitting PCPs of pseudo-H.

|  | **Cluster Method** GGA | **Pseudo-molecules** GGA | **Pseudo-molecules** HSE | **Previous works** GGA |
|---|---|---|---|---|
| **ZnO-(0001)** | 147.7 | 147.9 | 165.5 | 149.2[a] |
| **ZnO-(000$\overline{1}$)** | 63.1 | 66.7 | 113.9 | 59.9[a] |
| **ZnO-(11$\overline{2}$0)** | 51.7 (55.8) | 53.0 |  | 58.0[b] |
| **ZnO-(10$\overline{1}$0)** | 51.0 (53.9) | 54.0 |  | 61.2[b] |
| **GaN-(0001)** | 168.3 | 169.5 | 201.5 |  |

| | | | |
|---|---|---|---|
| GaN-(000$\bar{1}$) | 198.2 | 202.1 | 265.6 |
| GaN-(11$\bar{2}$0) | 103.0 (104.5) | 106.0 | |
| GaN-(10$\bar{1}$0) | 96.9 (98.0) | 102.1 | |

a, reference [31], original values are in the unit of J/m$^2$.

b, reference [52], original values are in the unit of J/m$^2$.

Using the PCPs of $H_{Zn}$, $H_O$, $H_N$, and $H_{Ga}$ obtained from both cluster method and pseudo-molecules, we obtained the surface energies of different surfaces, as listed in **Table 3**. All the calculations were performed at anion-rich conditions. Under such conditions, for ZnO, the surface energy of Zn-terminated (0001) surface is much higher than that of O-terminated (000$\bar{1}$) surface, while for GaN, on the contrary, N-terminated (000$\bar{1}$) surface has higher surface energy. The nonpolar surface energy of ZnO is similar with (000$\bar{1}$) surface energy (only slightly higher), while the nonpolar surface energy of GaN is much lower than that of the polar surfaces. From the differences between the sum of PCPs from polar surfaces and that from non-polar surfaces, we can conclude that, for the cluster method, the upper limit of the errors of the calculated surface energies for WZ polar surfaces should be within 4.1 meV/Å$^2$ for ZnO and 1.5 meV/Å$^2$ for GaN. We also applied the pseudo-molecule method for hybrid functional calculations, and found that the GGA functional will underestimate surface energies, but the values are still reasonable. Comparing our ZnO results with early works, we found that early works had comparable results with little differences. For GaN, to our best knowledge, there is no GGA result to make a comparison, but there were results for the reconstructed polar surfaces and non-polar surfaces based on hybrid functional calculations [30]. We noticed that the results in ref [30] are similar with our results, but our results exhibit higher self-consistencies. The large differences between GGA and HSE are due to the underestimation of band-gap for GGA functional. It can also be seen from **Table 3** that when we apply the pseudo-H passivation method to atoms with strong electronegative elements like O and N, the simple pseudo-molecule approach is good enough to give very accurate results for PCPs of anion. This implies that if the bottom surfaces of the slabs are terminated by anions with strong electronegativity, the pseudo-molecule approach already yields accurate results while saving much computing time.

Generally, the data in **Table 3** show that the surface energies of GaN are much higher than those of ZnO, suggesting that ZnO tends to wet the GaN substrate, while GaN is unlikely to wet ZnO. Therefore, it's very challenging to grow high quality GaN thin films on ZnO substrates, however, high quality ZnO thin film on GaN substrate is possible [53-65]. Indeed, 3D-like growth of GaN on ZnO substrate has already been observed [58,60]. On the other hand, on top of a single crystal GaN substrate, ZnO crystal films of high quality and sharp interfaces have been observed [63,64]. Thus our results are consistent with the experimental observations.

## IV. Conclusion

Although surfaces without reconstruction may not exist in experiments, the surface energies of such ideal surfaces are still important in studying reconstructed surfaces. In previous researches of the surface stability and growth kinetics [66-68], the ideal unreconstructed surfaces are often used as references for reconstructed surfaces. Therefore, we believe that our results may serve as reliable foundations in these studies.

In summary, we applied a novel and simple method to calculate the absolute surface energies of WZ polar surfaces both for ZnO and GaN with GGA and HSE functional. The errors of surface energies for WZ $(0001)/(000\bar{1})$ polar surfaces are within 4.1 $meV/Å^2$ for ZnO and 1.5 $meV/Å^2$ for GaN. We also obtained the accurate absolute surface energies of ZnO and GaN polar surfaces for HSE and GGA, respectively, for the first time. These accurate values can serve as important references for further studies on the growth kinetics of ZnO and GaN. Also, this method is generally applicable to determine the surface energies of other important wurtzite materials such as CdS, CdTe, InN and AlN.

## Acknowledgements

Part of the computing resources was provided by the High Performance Cluster Computing Centre, Hong Kong Baptist University. This work was supported by the start-up funding and direct grant with the Project code of 4053134 at CUHK.